\documentclass[linenumbers,trackchanges,twocolumn]{aastex7}

\usepackage{amsmath}
\usepackage{hyperref}
\usepackage{xcolor}
\begin{document}

\title{Core Scouring Dynamics and Gravitational Wave Consequences: Constraints on Supermassive Black Hole Binary Hardening}

\author[orcid=0000-0002-4231-7802,sname='CJ']{C.J. Harris}
\affiliation{University of Michigan Department of Astronomy \& Astrophysics}
\email[show]{cordellh@umich.edu}  

\author[orcid=0000-0002-1146-0198, sname='Kayhan']{Kayhan Gültekin} 
\affiliation{University of Michigan Department of Astronomy \& Astrophysics}
\email{kayhan@umich.edu}

\author[orcid=0000-0002-2183-1087, sname='Laura']{Laura Blecha} 
\affiliation{University of Florida, Department of Physics}
\email{laura.blecha@nanograv.org}

\begin{abstract}
We present a multi-messenger investigation of supermassive black hole binary (SBHB) hardening by jointly comparing models to the observed stellar mass deficits of nearby core galaxies and to nanohertz gravitational-wave background measurements. Using merger trees from the IllustrisTNG cosmological simulations, we construct a synthetic population of core galaxies and predict their stellar mass deficits under binary evolution driven by dynamical friction, stellar scattering, and gravitational-wave emission, assuming circular orbits. We calibrate the efficiency of stellar scattering by matching the observed relation between stellar mass deficit and galaxy stellar mass. The inferred hardening rates reproduce the observed core population when stellar scattering is more efficient than in our baseline prescription, consistent with additional environmental or dynamical effects. However, these rates do not produce the level of low-frequency attenuation required to match the apparent turnover in the nanohertz gravitational-wave background. This tension suggests that either additional physics—such as eccentricity evolution, gas dynamics, recoils, triple interactions, or changes in the black hole--galaxy scaling relation—or contributions from galaxy populations not represented by local core galaxies (e.g., massive power-law ellipticals) may be important. A comprehensive treatment of these effects is required before identifying the dominant mechanisms shaping the gravitational-wave spectrum.
\end{abstract}


\section{Introduction}
\label{sec:intro}
It is now an observational fact that all massive galaxies contain at least one supermassive black hole (SBH) at or near their centers \citep{1998AJ....115.2285M,1998Natur.395A..14R,2005SSRv..116..523F,2019ApJ...875L...1E}. Under $\Lambda\mathrm{CDM}$ cosmology, galaxies assemble by hierarchical merging, and observationally we see galaxies in various stages of the merging processes. It is inevitable that remnants of galactic mergers would then host multiple SBHs, which may eventually coalesce \citep{1980Natur.287..307B}. In the final stages of the coalescence the SBHs will produce strong nanohertz gravitational wave (GW) emission. The sum of all such emission should be detectable by pulsar timing arrays (PTAs) by measuring coherent deviations in the times of arrival of pulsar signals. Indeed, various collaborations using PTAs including the Parkes Pulsar Timing Array \citep{2023ApJ...951L...6R},  European Pulsar Timing Array \citep{2023A&A...678A..50E}, North American Nanohertz Observatory for Gravitational Waves (NANOGrav) \citep{2023ApJ...951L...8A}, and Chinese Pulsar Timing Array \citep{2023RAA....23g5024X} have confirmed a common red noise process seen in measurements of timing residuals and shown evidence for the Hellings \& Downs correlations \citep{1983ApJ...265L..39H}, the telltale sign of gravitational waves.

Using the NANOGrav 15 yr data set, \citet{2023ApJ...952L..37A} inferred the properties of a cosmic population of SBH binaries capable of producing the observed GWB, though any astrophysical interpretation based on gravitational wave data alone will carry degeneracies. For example, when explaining the higher than expected GWB amplitude it is difficult to constrain the contributions from an increased number density of SBHs versus that from an increased amplitude of SBH scaling relations. The degree to which dynamical friction versus stellar scattering and gas dynamics drive binary coalescence is another such degeneracy. These degeneracies can be broken by direct electromagnetic observations of the effects SBHs have on their local and/or galactic-scale environments. 

Examples of electromagnetic observations of pairs of SBHs are abundant in the literature. When the SBHs within a merger remnant are gravitationally unbound, they are referred to as a \emph{dual} SBH, and once they are bound they become an SBH \emph{binary} (SBHB). Many instances of the dual SBHs are known in the form of dual active galactic nuclei (AGN) \citep{2012AJ....143..119I,2016MNRAS.456.1595M,2019MNRAS.483.4242L,2019NewAR..8601525D, 2019ApJ...877...17F, 2020ApJ...892...29F, 2021ApJ...907...71F,2020ApJ...899..154S, 2021ApJ...922...83T, 2025ApJ...993..151B}, which are the natural precursors to SBHBs. The study of these systems can therefore provide crucial insights into SBHB demographics. Examples of this include the discovery of an inverse relationship between SBH separation and dual AGN activity and the fact that dual AGN mass ratios are always close to unity \citep{2016MNRAS.458.1013S,2016MNRAS.460.2979V}. Results such as these place important constraints on the overall population of SBHBs.

Currently, there is one confirmed SBHB \citep{2006ApJ...646...49R}, though many candidates exist \citep{2012ApJS..201...23E,2015ApJS..221....7R,2016MNRAS.463.2145C,2017MNRAS.468.1683R,2019NewAR..8601525D,2025arXiv250506221M, 2025arXiv250816534A}. Whether this lack of observed systems is because of predominately short/long merging timescales, or observational obstacles, is an open question.

Alongside direct detections of dual AGN and SBHBs, one can measure the effect a past binary has had on the host galaxy's structure. An SBH will sink to the bottom of the merger product's potential well under three main mechanisms \citep{1980Natur.287..307B}: (1) Dynamical friction, (2) loss-cone scattering, and (3) gravitational wave (GW) emission. It is in the dynamical friction regime of gas-rich remnants that dual AGN form. During the loss-cone (stellar) scattering phase, nuclear stars whose orbits cross within a few times the SBHB's semi-major axis tend to be ejected via three-body interactions, leading to an overall shrinking (hardening) of the SBHB's orbit \citep{1996NewA....1...35Q,2006ApJ...651..392S,2013CQGra..30x4005M}. If there are a suitable number of stars available to interact with the binary, stellar scattering will harden the binary into the separation regime where GW emission will dominate the evolution, eventually resulting in coalescence.

\begin{figure}
    \centering
    \includegraphics[width=1.0\linewidth]{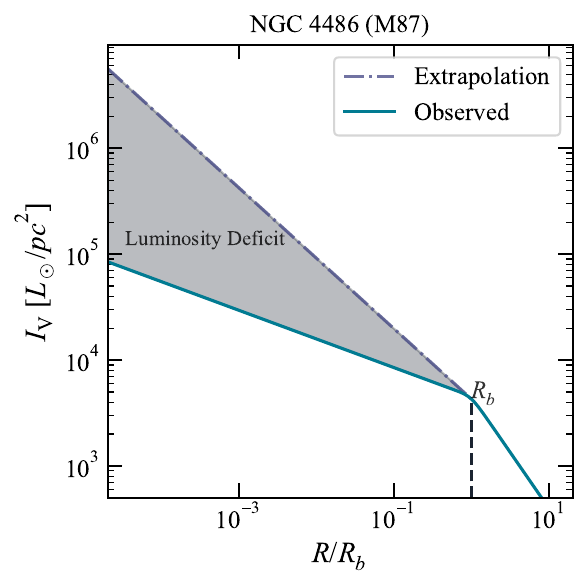}
    \caption{The luminosity deficit in core galaxies may be estimated by taking the integrated difference between a model of the pre-scoured profile and the observed surface brightness profile. Shown here is an example for the galaxy NGC 4486 (M87). Here, we model the surface brightness profile with the Nuker Law, shown in solid blue (see Eq.\  \ref{eq: Nuker}). Our estimate for the pre-scoured profile, shown in dashdot purple, is a power law with inner logarithmic slope equal to that at the break radius ($R_b$, see Eq.\ 
    \ref{eq: Plaw}). The break radius indicates the distance from the galaxy center where the profile transitions from the shallow inner region to the steep outer region and represents the size of the core.}
    \label{fig:Lumdef}
\end{figure}

The result of stellar scattering in a galactic nucleus is a core that is likely to persist indefinitely, appearing in observations as a luminosity deficit compared to the inward extrapolation of the outer light profile to small radii (Fig.\ref{fig:Lumdef}).  Nuclear star-formation is expected to erase cores in the case of a subsequent major, gas-rich merger \citep{1997AJ....114.1771F}, however. Indeed, the most massive early type galaxies (ETGs) are gas-poor and have surface brightness profiles which demonstrate luminosity deficits. These are referred to as ``cores", while their less luminous counterparts have brightness profiles that continue in a power-law fashion into the resolution limit. These latter ETGs dominate above $\mathcal{M}_V\sim-21$ \citep{1997AJ....114.1771F, 2006ApJS..164..334F,2007ApJ...664..226L, 2009ApJS..181..135H, 2009ApJS..181..486H} and are typically referred to as either ``cusps" or ``power-laws". 
The results of various N-body experiments alongside the observed, physical properties of core galaxies provide compelling evidence that SBHB activity is responsible for the shallow inner density profiles observed. Most massive ETGs have undergone at least one major merger since $z\sim 1$ \citep{2006ApJ...652..270B}, implying that at least one dual or binary SBH has operated in each of these galaxies. Between 50--70\% of the mass growth of massive ETGs has also occurred since $z\sim 1$, with about half of the growth caused by major mergers for galaxies $\lesssim10^{11} M_\odot$ \citep{2012A&A...548A...7L} while major mergers dominate mass assembly for the most massive ETGs \citep{2009ApJ...697.1369B}.

Direct observational comparisons of the properties of core and power-law galaxies support the picture of core formation through SBHB activity. Analyses of galaxy kinematics have shown that cores have boxy isophotes, indicative of systems supported by anisotropic velocity dispersions---the hallmark of gas-poor violent relaxation---while power-laws have disky isophotes characteristic of isotropic velocity dispersions and embedded disks \citep{1991A&A...244L..37N,1996ApJ...464L.119K,1997AJ....114.1771F, 2008gady.book.....B}. Consequently, cores are slow rotators and power-laws are fast \citep{1996ApJ...464L.119K, 2012ApJ...759...64L}, with gas-poor mergers tending to reduce the specific angular momentum of stellar components by $\sim30\%$, while gas-rich mergers increase angular momentum by $\sim10\%$ \citep{2018MNRAS.473.4956L, 2020MNRAS.494.5652W,2022ApJ...925..168Y}. Cores tend to be rounder than power-laws in their morphology \citep{1994AJ....108.1567J, 1994AJ....108.1598F, 2001MNRAS.326.1141R, 2005AJ....129.2138L}, while at the same time showing signs of triaxiality that core-less galaxies seem not to possess \citep{1996ApJ...464L.119K}.

The existence of correlations between the properties of a core-galaxy's black hole and core-galaxy properties and N-body simulation results further support the core scouring model. There is a tight correlation between the sphere of influence of a core galaxy's SBH and the size of the core \citep{2016Natur.532..340T, 2025arXiv251204178D} and many N-body experiments show that the mass ejected from the core scales as the mass of the binary and the number of dry mergers \citep[e.g.,][]{2006ApJ...648..976M,2012ApJ...749..147K, 2021ApJ...922...40D}. Indeed, using observed mass deficits in conjunction with the results of N-body experiments and cosmological magnetohydrodynamical simulations, \cite{2024MNRAS.528....1H} were able to show that it is possible to draw conclusions about the cosmic SBHB population from core galaxy properties.

Given the deep connection between core galaxy properties and SBHBs, the core galaxy population provides a promising avenue for studying the SBHB population and complements GWB studies, since cores are formed by the most massive SBHs during redshifts $0 < z \lesssim 2$, where most of the signal in the nanohertz GWB is generated \citep{2004ApJ...611..623S,2023ApJ...952L..37A}.

In this Paper we investigate how electromagnetic observations of the local core galaxy population can play a pivotal role in the astrophysical interpretation of the GWB. Specifically, we aim to place constraints on SBHB stellar hardening timescales---an important ingredient in astrophysical models of the GWB. We do this by comparing an observed population of core galaxies to a synthetic population. In section \ref{s: Methods} we describe the methods used to estimate stellar mass deficits and stellar masses from an observed sample of ETGs. Section \ref{s: results} contains the results of our statistical comparison between synthetic data sets and the observed sample. In section \ref{s:discuss} we discuss the caveats of the study and section \ref{s: sum} provides an overall summary.

Throughout we adopt $\Omega_M=0.3$, $\Omega_\Lambda=0.7$, $H_0=70 \ \mathrm{km\ s^{-1}\ Mpc^{-1}}$ and ensure all values/observations conform to this cosmology. Additionally, we adopt a \cite{2003PASP..115..763C} initial mass function (IMF) and all magnitudes used are in the Vega system.
\section{Methods}
\label{s: Methods}
Our goal is to make a connection between the observed light profiles of core galaxies and the SBHB population responsible for generating the GWB in the service of breaking degeneracies in the astrophysical interpretations of the GWB. We expect such a connection since core galaxies host the most massive black holes and are the results of major mergers, implying they were hosts of the most massive binaries in the past. Our approach to making this connection consists of (1) estimating the stellar mass deficits of a sample of observed core galaxies (Section \ref{ss: obs}) and (2) generating a model for the formation of galaxy cores using the results of N-body experiments and cosmological magnetohydrodynamical simulations (Sections \ref{ss: TNG} \& \ref{ss: core scouring model}).
\subsection{Observed Data \& Photometric Analysis}
\label{ss: obs}
Our observational dataset consists of 116 ETGs with Hubble Space Telescope (HST) observations and derived Nuker surface brightness profiles. We require HST observations, as the upper limit for the angular size of a galaxy core in the local Universe is on the order of a few arcseconds \citep{1997AJ....114.1771F}. There are two models used most frequently to describe the surface brightness of core galaxies are the  Core-Sersíc \citep{2003AJ....125.2951G} and Nuker \citep{1995AJ....110.2622L} profiles. There have been extensive discussions in the literature about the effectiveness of each and to what extent their core radius parameters can be said to be physically meaningful \citep[e.g.,][]{2003AJ....125.2951G,2006ApJS..164..334F, 2007ApJ...664..226L, 2009ApJS..181..486H}. In this work, we adopt the Nuker law, as the fits to the inner region are less sensitive to the global light profile. The functional form is given by
\begin{equation}
    I(R)=2^{\frac{(\beta-\gamma)}{\alpha}}I_b\left(\frac{R}{R_b}\right)^{-\gamma}\left[1+\left(\frac{R}{R_b}\right)^{\alpha}\right]^{\frac{(\gamma-\beta)}{\alpha}},
    \label{eq: Nuker}
\end{equation}
where $R$ is the projected radial coordinate and $R_b$ is the core or `break' radius, indicating the transition between the inner and outer profile. The parameter $\gamma$ sets the logarithmic slope of the profile as $R\rightarrow0$, while $\beta$ sets the slope as $R\rightarrow\infty$, with $\alpha$ controlling the sharpness of the transition between regimes. Lastly, $I_b=I(R_b)$ is the intensity at the break radius.

Photometric data for the galaxies in the sample were obtained from various sources. WFPC1 and WFPC2 $V$-band observations were obtained by \cite{1995AJ....110.2622L,1997AJ....114.1771F,2005AJ....129.2138L}. $R$-band WFPC2 photometry come from \cite{2001AJ....121.2431R}, while NICMOS $H$-band photometry were taken by \cite{2001AJ....122..653R} and \cite{2000ApJS..128...85Q}. Finally, \cite{2003AJ....125..478L} obtained WFPC2 $I$-band photometry for the BCGs in the sample. \citet{2007ApJ...664..226L} have transformed all parameters to $V$--band values, which we utilize here.

Nuker parameters are obtained from \cite{2007ApJ...664..226L} while mass-to-light ratios come from \cite{2009ApJS..181..486H}, who used the \cite{2003ApJS..149..289B} $\Upsilon_X$--$\mathrm{color}$ relation, corrected for the Chabrier IMF. We follow the same procedure to obtain mass-to-light ratios for galaxies that are not common between \cite{2007ApJ...664..226L} and \cite{2009ApJS..181..486H} but have Nuker parameters.

We estimate the stellar mass deficit of a core galaxy with
\begin{equation}
    M_\mathrm{def}=2\pi\Upsilon\int^{R_b}_0[I(R)-I_0(R)]R\ dR,
    \label{eq: mdef int}
\end{equation}
where $I(R)$ is the fitted parametric model of the observed surface brightness profile, $I_0(R)$ is a model of the original light profile (discussed in more detail below), and $\Upsilon$ is the galaxy's stellar mass-to-light ratio (Fig. \ref{fig:Lumdef}).

We take the limit of integration to be the break radius, which sets the size of the core and corresponds to the location of the maximum second derivative of the brightness profile in logarithmic coordinates. Based on direct comparisons to derivatives of observed light profiles $R_b$ carries with it a $\sim 30\%$ uncertainty \citep{2007ApJ...664..226L}.

Further, the break radius has a tight log-linear relationship with the sphere of influence of the SBH \citep{2016Natur.532..340T}. In the context of core scouring theory this is the gravitational reach of the binary that previously operated in the galactic nucleus. Beyond $R_b$ we expect the stellar population to remain relatively undisturbed.

\begin{figure*}
    \centering
    \includegraphics[width=1.0\linewidth]{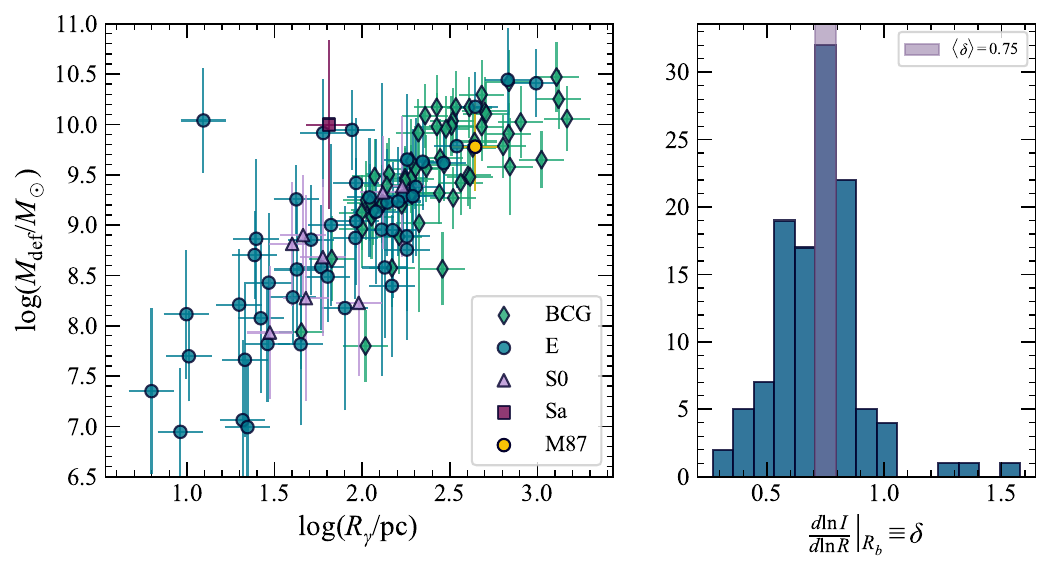}
    \caption{\emph{Left panel:} Mass deficit versus cusp radius for the observed sample of galaxies. The relation is clearly log-linear, with perhaps a weak break at the largest radii. Green diamonds show brightest cluster galaxies, blue circles indicate ellipticals, triangles indicate lenticulars, and the red square is a sole early-type spiral. For reference, M87 is represented by a yellow circle. \emph{Right panel:} The distribution of logarithmic slopes $\delta$ used to compute the stellar mass deficits, with the peak located at 0.75. This agrees well with the range of 0.7--0.8 \cite{2009ApJS..181..486H} found for progenitors of core galaxies.}
    \label{fig:Obs sample}
\end{figure*}

With the physical significance of $R_b$ in mind, we take as our model of the original inner light profile a power-law with a logarithmic slope equal to that of the observed surface brightness profile at the break radius $R_b$:
\begin{equation}
    I_0(R)=I_b\left(\frac{R}{R_b}\right)^{-\delta} \ ; \ R\in (0,R_b],
    \label{eq: Plaw}
\end{equation}
where $\delta=(\beta+\gamma)/2$. We find the peak of the distribution of slopes to fall comfortably within the range expected of pre-merger nuclear profiles modeled with the Nuker law \citep[][see Fig.\ \ref{fig:Obs sample}]{2009ApJS..181..486H}.

Inserting equations \ref{eq: Nuker} and \ref{eq: Plaw} into Eq.\ \ref{eq: mdef int} and performing the integration yields
\begin{multline}
    M_\mathrm{def}=2\pi \Upsilon_V I_bR_b^2 \\ \times\ \left[\frac{1}{2-\delta} +\frac{2{}_2F_1\left(1,\frac{2+\alpha-\beta}{\alpha},\frac{2+\alpha-\gamma}{\alpha},-1\right)}{\gamma-2}\right],
    \label{eq: Mdef}
\end{multline}
where $_2F_1(a, b, c, d)$ is the Gauss hypergeometric function. Figure \ref{fig:Obs sample} shows the resulting mass deficits plotted against the ``cusp radius"
\begin{equation}
    R_\gamma\equiv R_b\left(\frac{\gamma'-\gamma}{\beta - \gamma'}\right)^{1/\alpha}.
\end{equation}
The cusp radius is a representative scale that is more tightly correlated with other galaxy properties than is $R_b$. This is largely caused by the fact that $\beta$ is correlated with the residuals of the break radius--luminosity ($R_b$--$L$) relation \citep{2007ApJ...662..808L}. The correlation arises from the fact that the range of $\gamma$ values for core galaxies is smaller than the range of $\beta$ values available. Physically, the cusp radius is the radius at which the negative logarithmic slope of a galaxy's surface brightness reaches a prespecified value $\gamma'$. We set $\gamma'=1/2$ as suggested by \cite{1997ApJ...481..710C} and $\cite{2007ApJ...662..808L}$.
\subsection{Illustris TNG300 \& TNG-Cluster Sample}
\label{ss: TNG}
Here we use IllustrisTNG \citep{2018MNRAS.475..676S} as the base of our framework to simulate the effects of SBHB hardening timescales on the prevalence and properties of scoured cores. The IllustrisTNG project is a cosmological magnetohydrodynamical simulation suite for galaxy formation and evolution, spanning a range of volume and resolution. A major component of the simulation suite is the \texttt{sublink} algorithm \citep{2015MNRAS.449...49R} used to construct subhalo merger trees, where ``subhalo" is interchangeable with ``galaxy". We use the results of the publicly available TNG300-1 and TNG-Cluster simulations, as described in \cite{2019ComAC...6....2N} and \cite{2024A&A...686A.157N}, to create a population of synthetic core galaxies with known mass deficits which we then compare to observed core galaxies. We selected the 300 $\mathrm{Mpc}$ volume with supplementary TNG-Cluster subhalos to provide us with the largest population of massive galaxies possible and to span the range of galaxy masses in the observed population of cores. To this end, we include all subhalos with stellar masses above $10^{10} \ M_\odot$ in our analysis.

The IllustrisTNG suite does not possess the resolution scale required to resolve core structure, however. TNG300-1, the highest resolution of the largest volume simulations, has a softening length for collisionless particles of $\epsilon_\mathrm{DM,\star}=1.4 \ \mathrm{kpc}$, whereas cores are $\lesssim 1 \ \mathrm{kpc}$ in extent. Accordingly, we must perform post-processing prescriptions to establish mass deficits for TNG300 and TNG-Cluster subhalos. To do this, we use the results of N-body scattering experiments and potential--density pair models alongside \texttt{sublink} merger trees to calculate the expected mass deficit for each subhalo at $z=0$.
\subsection{Core Scouring Model}
\label{ss: core scouring model}
Galaxy mergers are highly non-linear processes, altering the mass and stellar density profiles as well as injecting mass into the nucleus via merger induced nuclear starbursts. All of these phenomena need to be modeled if we are to track the evolution of a SBH pair. In our post-processing model of core scouring, we use observation- and simulation-informed prescriptions to determine whether core scouring occurs within a given galaxy hosting an SBH pair, and how much mass is scoured out. We consider galaxy specific star-formation rate ($\mathrm{sSFR})$ to determine if any core that would have formed would be erased. We use analytically well-behaved prescriptions---with observationally informed scaling relations---for galaxy density profiles and bulge fractions. We additionally use the results of galaxy merger simulations to estimate the amount of mass injected into a galactic nucleus, as this will alter the stellar density profile and thus the background of stars available to interact with an SBH pair. Once the environment is determined we consider the SBHs and their hardening mechanisms.
\begin{figure}
    \centering
    \includegraphics[width=1.0\linewidth]{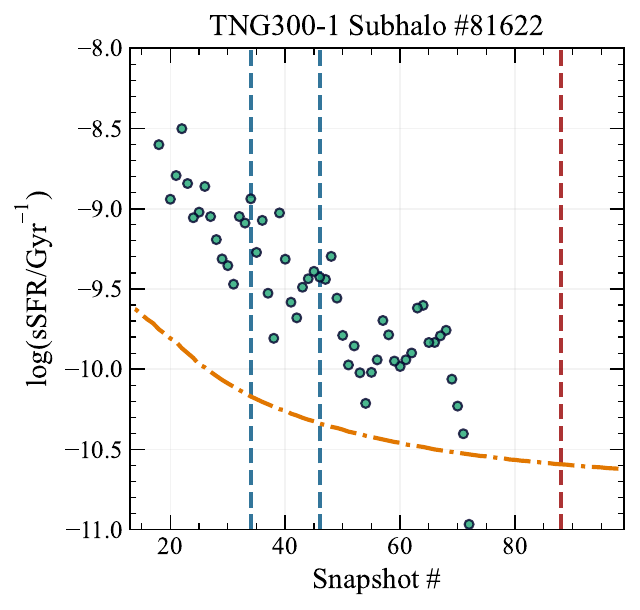}
    \caption{Shown here is the specific star-formation rate and mass history of a TNG300-1 subhalo in our sample. The dashed lines show the snapshots in the simulation where the subhalo underwent a major merger (i.e., $q_\mathrm{gal}\ge1/3$). For our core scouring model we label a merger gas-rich (wet) if an episode of star-formation occurs during or any time after the merger. Conversely, if during and after a merger $\mathrm{sSFR} < 1/3t_H$ the merger is labeld gas-poor (dry). Here $t_H$ is the Hubble time (orange dash-dot curve) at the corresponding redshift. It is only the dry mergers in our model that contribute to core scouring. For this example subhalo, only the most recent major merger is considered dry (red dashed line). The preceding mergers induced bursts of star formation, and count as wet mergers (blue dashed lines). Therefore, the wet merger at snapshot 46 would be the first merger considered in the analysis of this subhalo, used to set the inner logarithmic density slope $\lambda_p$ (\ref{eq:plaw slope}). Core scouring would then take place during the subsequent dry merger at snapshot 88, altering the shape of the nuclear profile.}
    \label{fig:sfrhist}
\end{figure}
\begin{figure*}[htbp]
    \centering
    \includegraphics[width=1.0\textwidth]{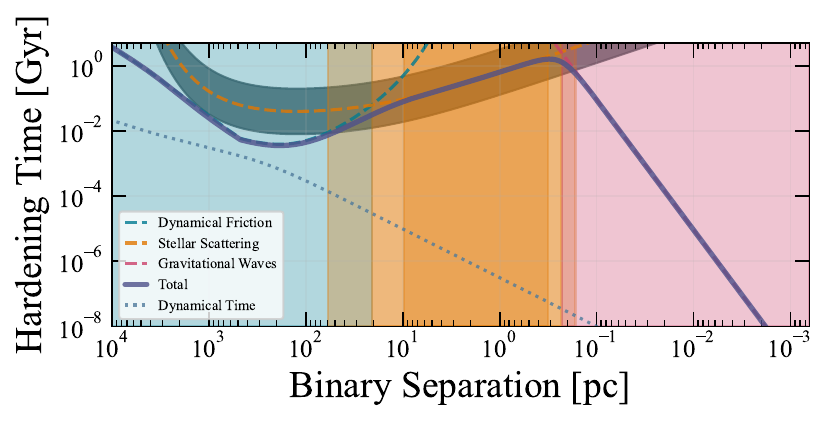}
    \caption{The plot shows hardening time in years on the y-axis and binary separation in parsecs on the x-axis, decreasing from left to right. The solid purple line shows the total evolutionary track for a representative binary in our model, which is a sum of the dynamical friction (blue dashed line), loss-cone scattering (orange dashed line), and gravitational wave emission (pink dashed line) hardening processes. The dotted line is the dynamical time $\tau_\mathrm{dyn}=r/v_c$ of the host galaxy, for reference. The dark shaded region around the loss-cone scattering curve denotes a range of possible $\hat H$ values from $0.1$--$5.0$. The colored shaded regions correspond to the separations where each of the three hardening mechanisms dominates. The opaque orange region shows how inefficient hardening (small values of $\hat H$) leads to a smaller range of separations for which stellar scattering dominates the hardening rate, whereas efficient hardening leads to a broader range of separations, shown by the transparent orange. The boundaries of these shaded regions come from the intersection of the lower and upper bounds of the stellar-scattering curves with the dynamical friction and gravitational wave curves. Importantly, the efficiency parameter controls the separation at which gravitational waves dominate angular momentum loss, ultimately impacting the shape of the GWB.}
    \label{fig:binary-evolution}
\end{figure*}
For each subhalo in our sample we obtained the \texttt{sublink} merger tree data, selecting for mergers with galaxy mass ratios $q\ge1/3$ (i.e., major mergers). Since core galaxies are the result of gas-poor (`dry') mergers, it is necessary to characterize the gas content of the merging progenitors. We do this by examining the star-formation history of each subhalo alongside the snapshot where each major merger occurred. A merger is labeled gas-rich (`wet') if the $\mathrm{sSFR}$ during or any point after the merger exceeds $1/3t_H$ where $t_H$ is the Hubble time at the corresponding redshift. Otherwise, the merger is labeled dry (see Figure \ref{fig:sfrhist}).

Since core galaxies are the results of mergers between power-law galaxies, which themselves were the result of gas-rich mergers between disky galaxies, we take the starting point for our model to be the most recent wet, major merger a given subhalo has undergone. If the subhalo has not undergone any wet mergers, we instead begin with the earliest dry merger. We model the stellar density of the host galaxy bulge with a Dehnen profile \citep{1993MNRAS.265..250D}
\begin{equation}
    \rho_\star(M,\lambda)=\frac{(3-\lambda)M}{4\pi}\frac{r_c}{r^\lambda(r+r_c)^{4-\lambda}},
    \label{eq: dehnen}
\end{equation}
with $r_c(M,\lambda)=(2^{1/(3-\lambda)}-1) \ r_{1/2}(M)$. We follow \cite{2017MNRAS.470.1738C} and use the $M$\textendash$ r_{1/2}$ relation \citep{2008MNRAS.386..864D}: 
\begin{equation}
    r_{1/2}(M)=239\left(\frac{M}{10^9 \ M_\odot}\right)^{0.569} \ \mathrm{[pc]}\ \ \pm \ 0.2 \mathrm{\ dex}.
\end{equation} The inner logarithmic slope $\lambda$ sets the density within the characteristic radius and is a key parameter in the evolution of each SBHB, the treatment of which we describe below. 

The bulge mass of the remnant is determined by assigning a constant bulge fraction $B/T=0.615$ to each subhalo based on observations of local ETGs \citep{2014ApJ...788...11L, 2014MNRAS.441..599B}.

We determine the value of $\lambda$ in the density profile based on the inferred dissipational component (i.e., the fraction of stars in the nucleus formed as the result of a merger-induced starburst) of the gas-rich merger:
\begin{equation}
    \langle f_\mathrm{inj}\rangle=\left[1+\left(\frac{M_\mathrm{bulge}}{M_0}\right)^{\nu}\right]^{-1}.
\end{equation}
 \cite{2009ApJS..181..486H} find that, at $z=0$, core ellipticals follow $(M_0, \nu)=(10^{8.8}, 0.35)$, while the best-fit values for cusp ellipticals are $(M_0, \nu)=(10^{9.2}, 0.43)$, where $M_0$ is in solar masses. We fix these values to those of the cusp ellipticals (with a constant factor of 2 intrinsic scatter at each mass) when determining the amount of stars that would have been injected into the core during the last merger induced starburst.

Thus, the amount of mass injected into the galactic nucleus is $M_\mathrm{inj}=\langle f_\mathrm{inj}\rangle M$. If we assume that the dissipational component is contained within the characteristic stellar radius $r_c$, then we can define the inner logarithmic density slope using the fact $M_\mathrm{inj}=4\pi \int_0^{r_c}(\Delta \rho)r^2dr$. Integrating and solving for the new inner logarithmic density slope, given a fixed $M_\mathrm{inj}$ yields
\begin{equation}
    \lambda_p=\frac{\log [2^{\lambda_0}(1-f_\mathrm{inj})+8f_\mathrm{inj}]}{\log2},
    \label{eq:plaw slope}
\end{equation}
where $\lambda_0=1.0$ is the initial inner logarithmic density slope of the primary progenitor. Note that if the first merger was gas-poor then $f_\mathrm{inj}=0$ and $\lambda_p=\lambda_0$.

We model the dark matter component with an NFW profile
\begin{equation}
    \rho_\mathrm{dm}(r)=\frac{\rho_s}{(r/r_s)(1+r/r_s)^2},
\end{equation}
where $r_s$ is the scale radius defined such that the scale density obeys $\rho_s=4 \ \rho_s(r_s)$. In general, the scale density is \begin{equation}
    \rho_s=\frac{200}{3}\rho_\mathrm{cr}(z)\frac{\mathcal{C}^3}{\mathrm{ln}(1+\mathcal{C})-\mathcal{C}/(1+\mathcal{C})},
\end{equation} 
where $\rho_\mathrm{cr}(z)=3H_0(z)^2/(8\pi G)$ is the critical density of the Universe at redshift $z$ and $\mathcal{C}$ is the concentration parameter (here we adopt the prescriptions of \citealt{2016MNRAS.457.4340K}).

In this work we aim to isolate the effects of stellar scattering and therefore do not explicitly treat SBHB interactions with gas. That is, we model a gas-free environment. We return to the importance of gas dynamics in the form of dynamical friction and circumbinary disk torques in our discussion (section \ref{s:discuss}) 

% Given that core galaxies are formed from dry mergers in the sense that the gas involved is not cold enough to form stars, we make the assumption that the gas is not dynamically relevant for the SBHBs and do not consider effects due to the formation of a circumbinary disk or gaseous dynamical friction.

The remnant galaxy is populated with an SBHB of mass $M_\mathrm{bin}=M_\mathrm{BH,1}+M_\mathrm{BH,2}$ and mass ratio $q=M_\mathrm{BH,2}/M_\mathrm{BH,1}$ where the black hole masses are determined from the primary and secondary progenitors of the merger via the $M$--$M_\mathrm{BH}$ relation \citep{2013ARA&A..51..511K}
\begin{equation}
    M_\mathrm{BH}=0.49\times10^9\left(\frac{M}{10^{11}M_\odot}\right)^{1.16} \pm0.28 \ \mathrm{dex}.
\end{equation}
The two SBHs are evolved within the density background of the remnant formed by the progenitors, beginning at a separation of $2r_{1/2}$. This background consists of the stellar and dark matter components.

Once the density background has been set, the semi-major axis of each SBHB evolves according to three physical effects: (1) Dynamical friction, (2) loss-cone (stellar) scattering, and (3) gravitational wave emission. The total evolution of the binary during these three stages determines how much stellar mass is removed from the core. Throughout we assume that binary orbits are initially circular and remain so until coalescence. We discuss the potential effects of lifting this assumption in section \ref{ss: ecc}.
\subsubsection{Dynamical Friction}
\label{sss: DF}
At large separations the semi-major axis evolution is driven primarily by dynamical friction of the secondary BH and its accompanying nuclear star cluster (NSC), if any, against the background stars and dark matter. The frictional force experienced is given by \citep{1943ApJ....97..255C}
\begin{equation}
    F_\mathrm{df}(a)=M_\mathrm{eff}\left.\frac{dv}{dt}\right|_\mathrm{df}=-\frac{2\pi^2G^2M_\mathrm{eff}^2\rho}{v_c^2}\mathrm{ln}\Lambda,
\end{equation}
where $v_c$ is the circular velocity at each separation, $\rho=\rho_\star+\rho_\mathrm{dm}$ is the sum of the stellar and dark matter density at each separation, $\ln\Lambda$ is the Coulomb Logarithm (set here to a constant value of $15$, following the results of \citealt{2017MNRAS.464.3131K}), and $M_\mathrm{eff}=M_\mathrm{BH,2}+M_\mathrm{NSC}(a)$ is the effective dynamical friction mass. Modeling a bare secondary SBH moving under dynamical friction has the potential to drastically underestimate the effective mass \citep{2017MNRAS.464.3131K}, thereby increasing the hardening timescale. This is particularly important at large separations, where dynamical friction is weak.

As the secondary BH and its NSC's orbit decays, tidal forces strip away stellar mass, eventually leaving the secondary bare. The tidal radius of the NSC, assuming it is described by a singular isothermal sphere in a circular orbit, is \citep{2008gady.book.....B}
\begin{equation}
    r_t = \sqrt{\frac{2\sigma_\star^2}{GM_h}}\left(3-\frac{d\mathrm{ln}M_h}{d\mathrm{ln}a}\right)^{-1/2}a^{3/2}
\end{equation}
with $\sigma_\star$ the one-dimensional stellar velocity dispersion of the NSC (taken to be that of the secondary BH's host galaxy) and $M_h=M_\star(a)+M_\mathrm{dm}(a)$ the cumulative mass of the remnant galaxy. The initial mass of the NSC is determined from the scaling relation \citep{2006ApJ...644L..21F, 2013ApJ...763...76S}
\begin{equation}
    \log(M_\mathrm{NSC})=2.11\log\left(\frac{\sigma_\star}{54\ \mathrm{km \ s}^{-1}}\right)+6.63,
\end{equation}
with an applied scatter of $0.55\ \mathrm{dex}$.

Applying conservation of momentum gives the semi-major access evolution as a function of time:
\begin{equation}
    \left.\frac{da}{dt}\right|_\mathrm{df}=-2\frac{\tau_\mathrm{dyn}}{\eta}\left(\frac{M_\mathrm{eff}}{({dM_\mathrm{eff}/}{da})a+M_\mathrm{eff}}\right)\left.\frac{dv}{dt}\right|_\mathrm{df}.
\end{equation}
In the above equation $\tau_\mathrm{dyn}=a/v_c$ is the dynamical time and $\eta$ is an attenuation factor. The attenuation factor is required because the dynamical friction evolution stated up until now is only valid outside separations where three-body loss-cone interactions are important. Thus, we apply a cutoff at $r_h=(M_\mathrm{bin}/M_\star)r_c$, the ``hard binary radius",  \citep{1980Natur.287..307B} such that $\eta=\exp\{r_h/a\}$. The blue curve in Figure \ref{fig:binary-evolution} shows an example of the dynamical friction hardening timescale for a representative binary in our model.
\subsubsection{Loss-cone Scattering}
\label{sss: loss cone}
\begin{figure}
    \centering
    \includegraphics[width=1.0\linewidth]{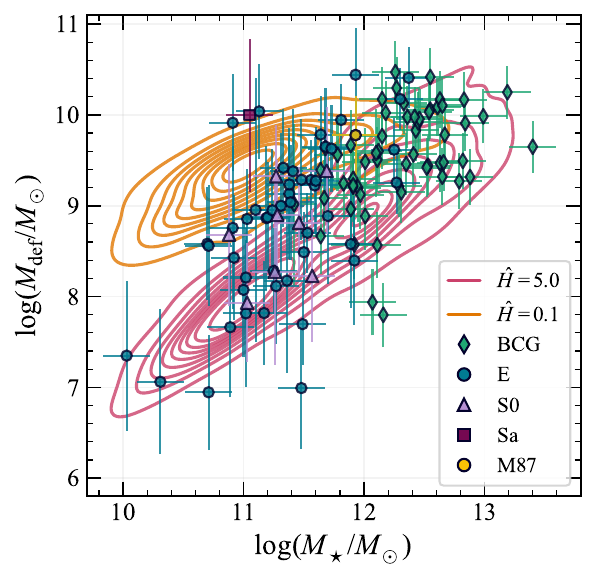}
    \caption{Mass deficit versus stellar mass for our sample of core galaxies alongside KDE contours generated using a high efficiency factor ($\hat H=5.0$) in pink and a low efficiency factor ($\hat H=0.1$) in orange. The KDE contours show that the efficiency of stellar hardening is inversely proportional to the amount of stellar mass ejected, as is required by equations \ref{dadt_lc} and \ref{eq: Mej}. The figure also shows that our model is sensitive to the density of points in the $M_\mathrm{def}$--$M_\star$ plane. Inefficient hardening leads to a denser clustering of points at lower masses not seen at higher efficiencies. The overall slope of the trend also depends sensitively on the hardening efficiency.}
    \label{fig:parmcomp}
\end{figure}

At separations $a\lesssim r_c$ the stellar density is sufficiently high that three-body interactions begin to dominate over dynamical friction. The evolution proceeds according to
\begin{equation}
    \left.\frac{da}{dt}\right|_\mathrm{lc}=-H(\hat H,a,q)\frac{G\rho_\star(r_\mathrm{soi})}{\sigma_\star}a^2\exp\{-a/r_c\},
    \label{dadt_lc}
\end{equation}
with $r_\mathrm{soi}$ the binary sphere of influence. Here we utilize the fits to N-body results given by \citet{2006ApJ...651..392S} along with an efficiency factor $\hat H$ to define the dimensionless hardening rate $H$. The variable $\hat H$ is the key parameter in our model. Values of $\hat H$ above unity describe more ``efficient" hardening during the stellar scattering regime, in the sense that the hardening timescale decreases. More efficient hardening during this stage of evolution implies binaries move faster through the stellar-scattering regime than the standard prescription, suggesting non-ejective mechanisms. On the other hand, values of $\hat H$ less than unity indicate ``inefficient" hardening. Sources of inefficiency could include problems with the assumption of a full loss-cone or with our choice of black hole mass scaling relation. See section \ref{s:discuss} for a more detailed discussion. The overall effects of varying $\hat H$ in our model are shown in Figure \ref{fig:parmcomp}. Importantly, $\hat H$ controls at which separations a given SBHB transitions from one evolutionary phase to the other (see Figure \ref{fig:binary-evolution}). More efficient stellar hardening will tend to cause a binary to enter the GW dominated regime at smaller separations, attenuating the low frequency signal as more orbital energy is transferred to interloping stars instead of being radiated as gravitational waves.

The preceding prescription for loss-cone scattering assumes that the stellar distribution is uniform and that the velocity distribution is Maxwellian. We set $\rho_\star=\rho_\star(r_\mathrm{soi})$ to be consistent with the assumption that the loss-cone is always full \citep{2002MNRAS.331..935Y,2006astro.ph..1520H, 2006ApJ...651..392S, 2012ApJ...749..147K, 2017MNRAS.464.2301G}. We also adopt an exponential turn-on at the characteristic stellar radius $r_c$. The orange curve in Figure \ref{fig:binary-evolution} shows an example of the loss-cone hardening timescale for a representative binary in our model.
\subsubsection{Gravitational Wave Emission}
\label{sss: gw}
Upon reaching separations below $\sim 1 \ \mathrm{pc}$ SBHB's become largely decoupled from their environment and dissipate orbital energy almost purely through GW emission.

For a circular binary evolving solely under the emission of gravitational radiation we have \citep{1964PhRv..136.1224P}
\begin{equation}
    \left.\frac{da}{dt}\right\vert_\mathrm{gw}=-\frac{64}{5}\frac{G^3}{c^5}\frac{M_\mathrm{bin}^3}{q(1+q)^2a^3}.
\end{equation}
Accordingly, binaries spend less time at small separations and consequently less time emitting at high frequencies. The amount of time SBHBs spend emitting GWs in a given frequency bin depends on the timing of the SBHB's decoupling from its environment. As mentioned in section \ref{sss: loss cone}, the efficiency of stellar scattering sets the decoupling time. The pink curve in Figure \ref{fig:binary-evolution} shows an example of the GW hardening timescale for a representative binary in our model.

\subsubsection{Stellar Mass Ejection}
\label{sss: mej}
As the SBHB interacts with stars whose orbits occupy the loss cone, energy will tend to be extracted from the binary and put into the orbits of the stars, ejecting them from the core. We calculate the total mass in stars ejected in this way as
\begin{equation}
    M_\mathrm{ej}=-M_\mathrm{bin}\int \frac{J(\hat H,a,q)}{a}\frac{\dot{a}\vert_\mathrm{lc}}{\dot{a}\vert_\mathrm{df} + \dot{a}\vert_\mathrm{lc}+\dot a\vert_\mathrm{gw}}da.
    \label{eq: Mej}
\end{equation}
The dimensionless ejection rate $J$ is the fit obtained by \cite{2006ApJ...651..392S}, modified by our efficiency parameter $\hat H$ such that 
\begin{equation}
    J=\int^\infty_0f(v,\sigma_\star)\frac{\sigma_\star}{v}4\pi v^2\ dv \int^\infty_0 \frac{F_{\rm ej}(x)}{H(\hat H,a,q)}\ dx,
\end{equation}
where $v$ is the speed of an interacting star, $x$ is the ratio of maximum to minimum impact parameters, $f(v,\sigma_\star)$ is the Maxwellian velocity distribution, and $F_{\rm ej}$ is the fraction of ejected stars. An efficient hardening rate ($\hat H > 1.0$) can therefore be interpreted as a decrease in the fraction of stars ejected by the binary, resulting in a lower stellar mass deficit. Since $dJ/dv=f(v)$, where $f(v)$ is the velocity distribution of ejected stars, this is also a statement that an efficient hardening rate decreases the fraction of ejected stars \emph{uniformly} across velocity bins in our model. This means that the shape of the distribution remains unmodified for varying values of $\hat H$. From an energetics perspective, our model is agnostic about where the binary orbital energy that would have been lost to hyper-velocity stars goes when $\hat H>1$. We discuss some of the possibilities in section \ref{s:discuss}.

This prescription carries the same assumptions as those mentioned in section \ref{sss: loss cone} with an additional assumption of a bulge-black hole mass relation of $M_\mathrm{BH}=0.0014M$, which sets the escape speed of the potential, modifying the mass ejection rate. The factor in equation \ref{eq: Mej} containing the hardening rates ensures that the mass ejection rate is attenuated while either dynamical friction or gravitational wave emission dominates the binary's evolution.
\begin{figure}
    \centering
    \includegraphics[width=1.0\linewidth]{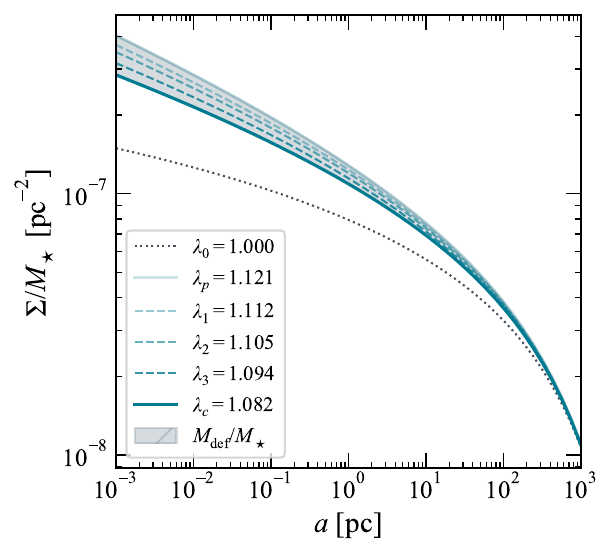}
    \caption{The figure shows an example of cumulative core scouring in our model. The y-axis shows the surface mass density of the remnant bulge divided by the stellar mass of the remnant, while the x-axis shows radial distance. Prior to the initial gas-rich merger, the primary progenitor is set to have a Hernquist profile ($\lambda_0=1.0$). The new, post-merger inner density slope $\lambda_p$ is then set using equation \ref{eq:plaw slope}. A subsequent gas-poor merger delivers a secondary black hole which forms a binary with the primary, scouring out a core ($\lambda_1$). Each successive dry merger results in shallower profiles each time ($\lambda_2$ and $\lambda_3$), culminating after the final merger in a profile with inner logarithmic slope $\lambda_c$. The shaded region represents the stellar mass deficit.}
    \label{fig:cumulative scouring}
\end{figure}
The stellar mass ejected from the core modifies the remnant's inner density profile. In reality, the size of the core $r_b$ is smaller than the characteristic stellar radius, but if we make the approximation that all stars are ejected from within $r_c$ and use $M_\mathrm{ej}=4\pi \int_0^{r_c}(\Delta \rho)r^2dr$ we find the new inner logarithmic slope to be
\begin{equation}
    \lambda_c=\frac{\log(2^{\lambda_p}-8f_\mathrm{ej})-\log(1-f_\mathrm{ej})}{\log2},
    \label{eq: lamc}
\end{equation}
where $\Delta \rho$ is the difference between the initial and post-scouring density profiles.

Since light profile shapes are conserved in gas-poor mergers \citep[e.g.,][]{2005MNRAS.362..184B}, $\lambda_c$ becomes $\lambda_0$ in the subsequent major merger (if there is one). Thus, in our model, core scouring is a cumulative process (see Fig. \ref{fig:cumulative scouring}) and the final mass deficit is $M_\mathrm{def}=\sum^\mathcal{N_\mathrm{dry}}_{i=1}M_\mathrm{ej,i}$ where $\mathcal{N}_\mathrm{dry}$ is the number of dry mergers contributing to core scouring.

In our analysis we only include systems for which the SBHB has coalesced before $z=0$ and therefore do not treat the effects of partial core scouring resulting from stalled binaries and do not include the effects of triple SBHs. We define the time to coalescence to be
\begin{equation}
    t_c=\int\left(\left.\frac{da}{dt}\right|_\mathrm{df}+\left.\frac{da}{dt}\right|_\mathrm{lc}+\left.\frac{da}{dt}\right|_\mathrm{gw}\right)^{-1}\ da
    \label{eq:tc}
\end{equation}
such that the binary evolution at any time is a superposition of all three effects.
\section{Results}
\label{s: results}
Here we describe the comparison of our core scouring model to the mass deficits observed in local core elliptical galaxies (section \ref{ss: compare}). We also investigate the implications our results have on astrophysical interpretations of the nanohertz GWB spectral strain amplitude (section \ref{ss: nano}).
\subsection{Comparison to the Observed Core Population}
\label{ss: compare}
\begin{figure}[htbp]
    \centering
    \includegraphics[width=1.0\linewidth]{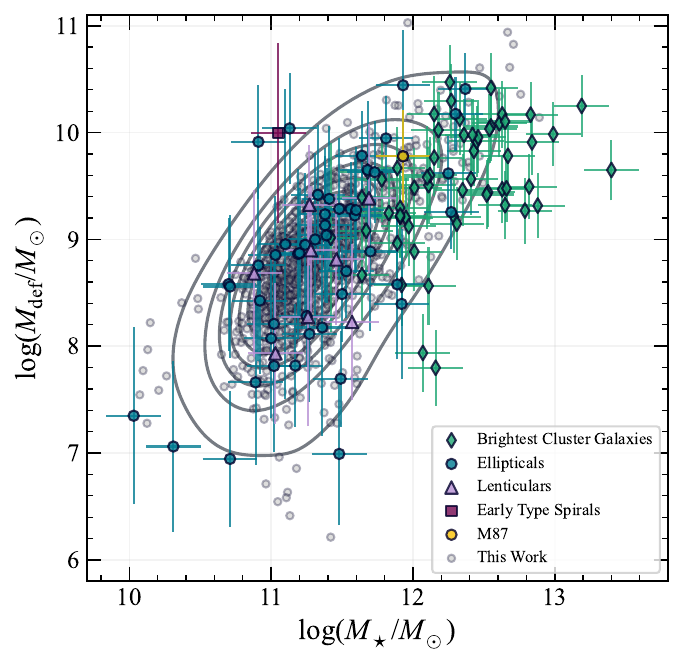}
    \caption{The figure shows stellar mass deficit as a function of stellar mass for both the observed set of core galaxies and a sample generated using our core scouring model with $\hat H=1.6$ and simulated observational uncertainties. Synthetic data-points are shown in grey with the corresponding KDE and are in agreement with the observed data. We reproduce the slope and amplitude of the mass deficit--stellar mass relation.}
    \label{fig:best-fit}
\end{figure}
\begin{figure}
    \centering
    \includegraphics[width=1.0\linewidth]{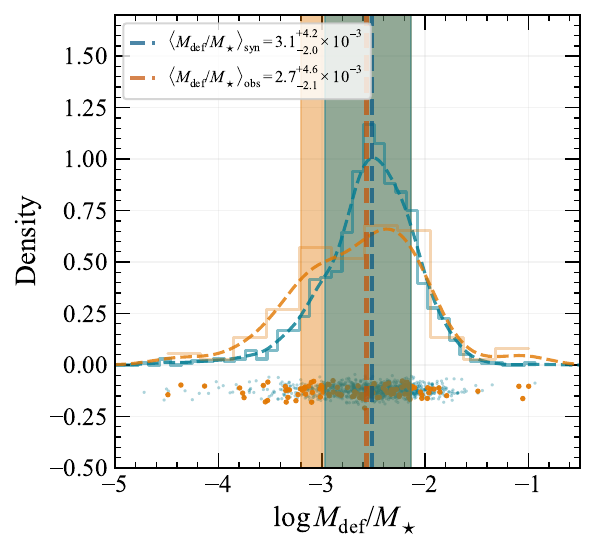}
    \caption{Plotted are the distributions of stellar mass deficit as a fraction of total galaxy stellar mass for both the best-fit synthetic core sample (blue) and the observed ETG sample (orange). Dashed lines indicate the median values of the distributions and shaded regions show the $1\sigma$ confidence intervals. While the observed, heterogeneous sample has a broader distribution than the synthetic sample, the medians agree. These values are consistent with other measured mass deficits \citep{2004ApJ...613L..33G, 2006ApJS..164..334F,2013ARA&A..51..511K,2013AJ....146..160R,2014MNRAS.444.2700D} and with those found using other core scouring models \citep[e.g.,][]{2006ApJ...648..976M,2007ApJ...662..808L,2009ApJS..181..135H,2012ApJ...749..147K, 2021ApJ...922...40D, 2024MNRAS.528....1H}.}
    \label{fig:fN}
\end{figure}
We use the Python Markov Chain Monte Carlo (MCMC) package \texttt{emcee} \citep{2013PASP..125..306F} to compare the model of cores outlined in section \ref{ss: core scouring model} to the observed mass deficits shown in section \ref{ss: obs}. We do this by constructing a likelihood function for a given data point $i$, with stellar mass and mass deficit values $M_{\star,i}, M_{\mathrm{def},i}$:
\begin{equation}
    \mathcal{L}_i(M_{\star,i}, M_{\mathrm{def},i}|\vec\theta)=K(M_{\star,i}, M_{\mathrm{def},i})
\end{equation}
where $K(M_{\star,i}, M_{\mathrm{def},i})$ is a kernel density estimation (KDE) evaluated at the data point using standard routines and $\vec\theta=[\hat H]$ in this case. The argument for the KDE is then a synthetic sample of core galaxies generated with a particular value of $\hat H$. Using MCMC methods alongside our log-likelihood ($\ln\mathcal{L}=\sum_i\ln\mathcal{L}_i$) and a flat prior on $\hat H$ of $[0.1, 5]$ we find the median value of the posterior samples and the 68\% confidence intervals to be $\hat H=1.64^{+0.16}_{-0.13}$. Thus stellar scattering appears to be more efficient, in the sense of a faster hardening rate and a shorter hardening time, than the N-body prescription by about a factor of 1.6. Figure \ref{fig:best-fit} shows a simulated population of core galaxies generated using the median value of the posterior $\hat H$ distribution along with the corresponding KDE. The observed sample of core galaxies is over-plotted for comparison. Figure \ref{fig:fN} shows distributions of the stellar mass deficit as a fraction of total galaxy stellar mass for both the observed sample and a synthetic sample derived from the best-fit model. While the observed, heterogeneous sample shows a much broader distribution than the synthetic sample, the median values are in agreement. We find for the observed sample $\langle M_\mathrm{def}/M_\star\rangle_\mathrm{obs}=2.7^{+4.6}_{-2.1}\times10^{-3}$ and for the best-fit model $\langle M_\mathrm{def}/M_\star\rangle_\mathrm{syn}=3.1^{+4.2}_{-2.0}\times10^{-3}$, where the uncertainties are given by the 68\% credible interval. These values are consistent with other measured mass deficits \citep{2004ApJ...613L..33G, 2006ApJS..164..334F,2013ARA&A..51..511K,2013AJ....146..160R,2014MNRAS.444.2700D} and with those found using other core scouring models \citep[e.g.,][]{2006ApJ...648..976M,2007ApJ...662..808L,2009ApJS..181..135H,2012ApJ...749..147K, 2021ApJ...922...40D, 2024MNRAS.528....1H}.

\begin{figure}
    \centering
    \includegraphics[width=1.0\linewidth]{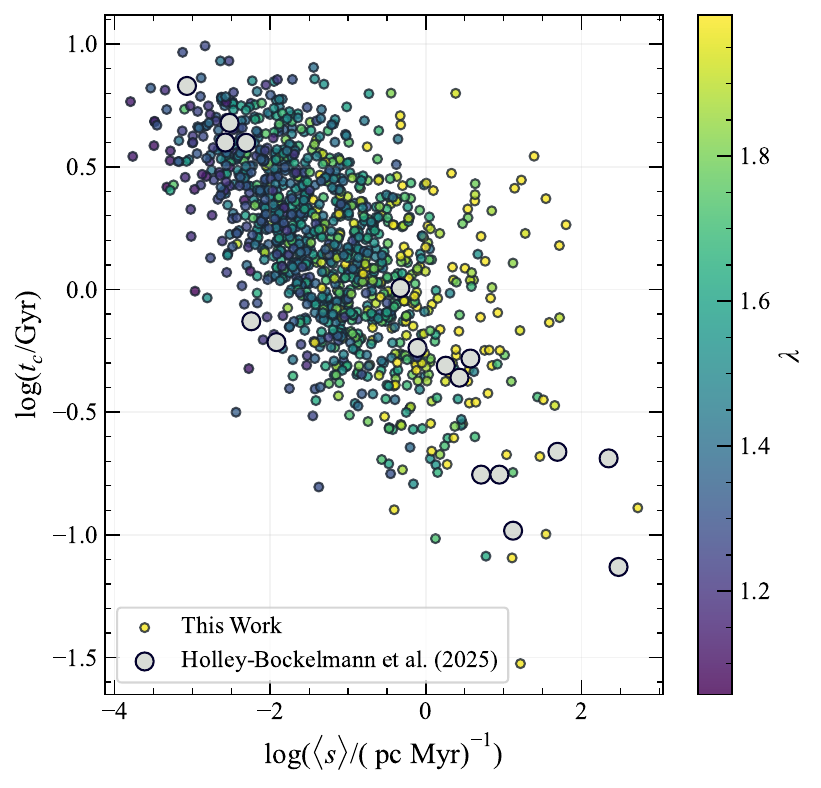}
    \caption{Coalescence time plotted against the average binary hardening parameter, $s=d(1/a)/dt$. Small circles show binaries from our best-fit semi-analytic model, with colors indicating the inner logarithmic slope of the post-scouring density profile, $\lambda$ (see Eq. 21), immediately after coalescence. Large circles show a subset of binaries from the high-resolution, gas-free, direct $N$-body simulations of \citet{2025arXiv250814253H}, which include non-zero initial eccentricities and prograde binary orbits relative to the host galaxy rotation. These simulations initialize binaries at separations of $\sim 10^2$--$10^3$ pc, substantially smaller than the typical initial separations of $\sim 10$ kpc in our model, and therefore yield systematically shorter total coalescence times. Since $t_c$ depends on the integral of the hardening history (including the dynamical friction stage) the agreement between the models, despite very different methods and initial conditions, suggests that in gas-poor major mergers the coalescence timescale is largely regulated by the stellar-hardening phase once the binary reaches the galactic center.}
    \label{fig:HB25}
\end{figure}

Further, we find agreement between the hardening timescales found here and the results of detailed N-body simulations of the galactic centers of a wide variety of galaxy types \citep{2025arXiv250814253H}. Figure \ref{fig:HB25} shows the binary coalescence time plotted against the average binary hardening parameter $s=d/dt(1/a)$ for both the binaries of our model fitted to the observed core population and a subset of N-body results presented by \citet{2025arXiv250814253H}. The binaries were evolved in prograde motion with respect to the host galaxy's rotation with non-zero eccentricity and an initial separation of $\sim10^2$--$10^3 \ \mathrm{pc}$---much closer than the typical value of $\sim10 \ \mathrm{kpc}$ in this work. As a result of the separation range explored in the N-body simulations the coalescence times are systematically shorter than those in our model. Regardless, we find our relationship between the coalescence times and the hardening parameter to be consistent with those found by \citet{2025arXiv250814253H}. The binaries of our model are colored according to the inner logarithmic slope of the density profile, showing that denser galactic nuclei are associated with faster coalescence times, as expected \citep{1996NewA....1...35Q, 2006ApJ...651..392S, 2012ApJ...749..147K}. The agreement between the models, despite very different methods and initial conditions, suggests that in gas-poor major mergers the coalescence timescale is largely regulated by the stellar-hardening phase, as opposed to the dynamical friction dominated regime.

\begin{figure}
    \centering
    \includegraphics[width=1.0\linewidth]{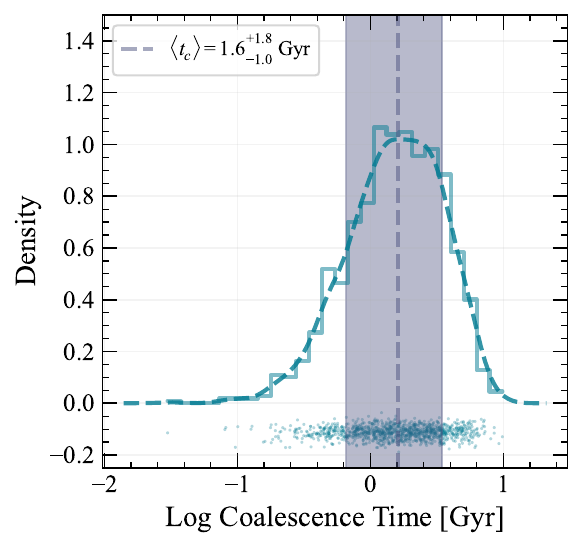}
    \caption{Here we plot the posterior distribution of coalescence times for the SBHBs in our best-fit model. We find the median value of the distribution to be $\langle t_c\rangle=1.6^{+1.8}_{-1.0}$, where the uncertainties are the 68\% confidence intervals. This is significantly larger than the \cite{2023ApJ...952L..37A} value of $\sim0.1 \ \mathrm{Gyr}$ and similar to timescales found by \cite{2017MNRAS.464.3131K} for massive binaries with high mass ratios.}
    \label{fig:merge_times}
\end{figure}

It is interesting to consider the distribution of coalescence times generated by our model when setting $\hat H$ to the median of the posterior distribution. We find that the median coalescence time for an SBHB in a gas-poor remnant is $1.6^{+1.8}_{-1.0}\ \mathrm{Gyr}$ where the uncertainties are the 68\% confidence intervals (see Fig. \ref{fig:merge_times}). If we instead consider the median of all evolutionary tracks in our model (Fig. \ref{fig:phenom-dadt-comp}) and compute the coalescence time we find $\langle t_c\rangle=1.4 \ \mathrm{Gyr}$, which closely corresponds to the peak in the overall distribution. This is significantly longer than the \cite{2023ApJ...952L..37A} value of $\sim 0.1 \ \mathrm{Gyr}$, but similar to timescales found by \cite{2017MNRAS.464.3131K} for massive binaries with high mass ratios.
Of further interest is to consider the overall effect of including NSCs during the dynamical friction phase of binary hardening. We find in our best-fit model that $\sim40\%$ of SBH pairs form with the secondary accompanied by an NSC. Significantly, when excluding NSCs from our model, we find no change in the posterior distribution of coalescence times, with $\langle\tau_c\rangle=1.8^{+2.1}_{-1.0}\ \mathrm{Gyr}$. We also find no significant change in the number of core-galaxies formed in the model. This result is due to the fact that the SBHs involved in core-galaxy formation are typically much more massive than NSCs, even when considering scatter. Thus we conclude that NSCs have a negligible impact on SBHB coalescence in core-galaxies.
\subsection{Implications for the Nanohertz Gravitational Wave Background}
\label{ss: nano}
In this section we investigate how the observed constraints on the stellar scattering hardening rate impact the astrophysical interpretation of the GWB.

The characteristic strain of the GWB caused by individual binaries over a logarithmic interval of frequency can be written as \citep{2001astro.ph..8028P, 2003ApJ...590..691W}
\begin{equation}
    h_c^2(f)=\int h_s^2(f_p)\frac{\partial^4 N}{\partial M_\mathrm{bin} \partial q\partial z \partial\ln f_p}dM_\mathrm{bin} dqdz.
    \label{eq: strain}
\end{equation}
In the above equation $h_s(f_p)$ is the GW spectral strain from a single binary as a function of its rest-frame orbital frequency. The derivative in the integrand is the number density of binaries per total mass, mass ratio, redshift, and logarithmic frequency bin. This can be recast in terms of the co-moving volumetric number density of binaries $n=dN/dV_c$ as \citep{2003ApJ...583..616J, 2008MNRAS.390..192S} 
\begin{multline}
    \frac{\partial^4 N}{\partial M_\mathrm{bin} \partial q\partial z \partial\ln f_p}=\\ \frac{\partial^3n}{\partial M_\mathrm{bin}\partial q \partial z}
    4\pi c\frac{f_p}{(df_p/dt)}(1+z)d_c^2,
    \label{eq:numden}
\end{multline}
where $d_c$ is the co-moving distance to a source at redshift $z$. Thus the GWB spectral strain depends on the duration a binary spends in a given logarithmic interval of frequency, and so on the details of binary evolution. This affords us a route to determine how our results impact astrophysical interpretations of the GWB.

 \cite{2023ApJ...952L..37A} were able to recover an SBHB model population by fitting against the derived GWB using models for the galaxy stellar mass function (GSMF), galaxy pair fraction (GPF), galaxy merger timescale (GMT), $M_\mathrm{BH}$--$M$ scaling relation, and the total binary hardening rate. Of interest here is the hardening prescription used in the study, referred to as the  ``phenomenological" hardening model. The model has functional form
\begin{equation}
    \left.\frac{da}{dt}\right|_\mathrm{ph}=H_a\left(\frac{a}{a_c}\right)^{1-\nu_\mathrm{in}}\left(1+\frac{a}{a_c}\right)^{\nu_\mathrm{in}-\nu_\mathrm{out}},
    \label{eq:phenom}
\end{equation}
and assumes that all binaries follow the same evolution.

\begin{figure}[htbp]
    \centering
    \includegraphics[width=1.0\linewidth]{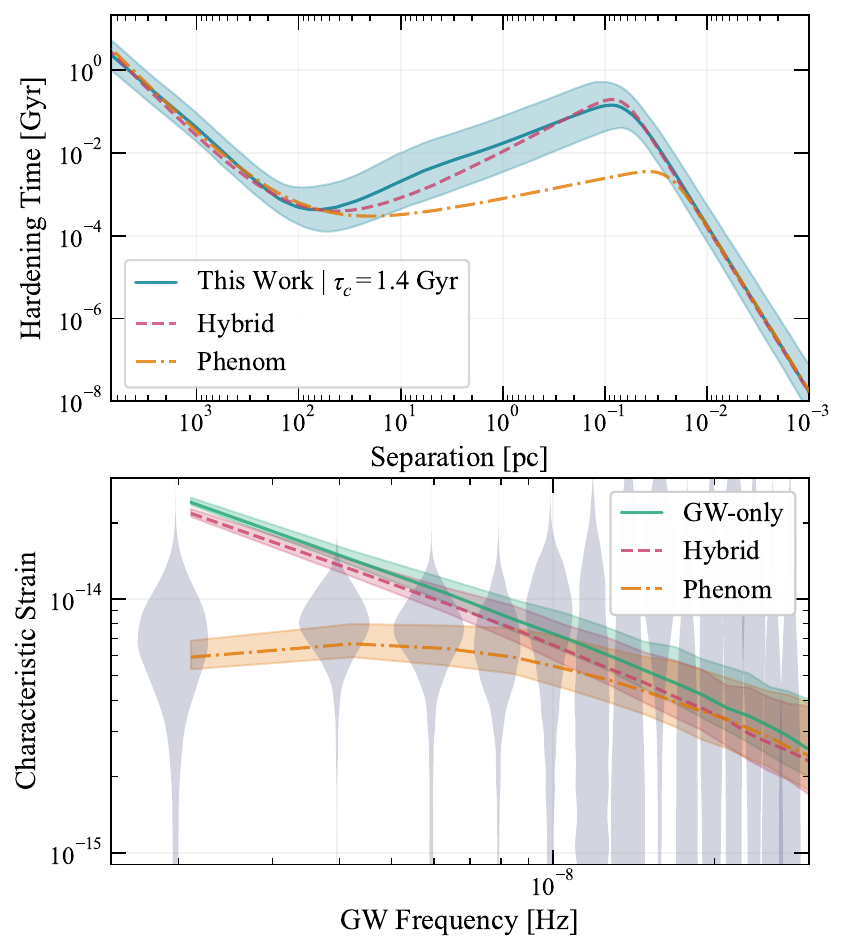}
    \caption{\emph{Top}: The plot shows the hardening timescale in gigayears versus binary separation in parsecs. The blue evolutionary track is the median hardening time across all binaries in the best-fit model (i.e., $\hat H=1.6$) with 68\% confidence intervals. The orange curve shows the \citep{2023ApJ...952L..37A} phenomenological hardening model with their best fit parameters plus hardening by gravitational waves, save the parameter $\tau_f$ which is set to match our $t_c$ for consistency. The pink track is the phenomenological model fit to the blue track.
    \emph{Bottom}: Shown on the y-axis is the characteristic strain amplitude of the GWB, with frequency in $\mathrm{Hz}$ on the x-axis. The gray violins show the GW spectrum derived from the NANOGrav HD-w/MP+DP+CURN models    \citep{2023ApJ...951L...8A, 2023PhRvD.108j3019L}.
    The colored curves are GW spectra resulting from SBHB populations generated using \texttt{holodeck}. Shown in green is the spectrum expected if the hardening mechanism that drove all binaries to coalescence was GW emission alone. The orange spectrum is the \cite{2023ApJ...952L..37A} fit to the GWB data. The pink hybrid model is comprised of the phenomenological model where the parameters in equation \ref{eq:phenom} are set by the results of our model. Shaded regions indicate the 68\% confidence intervals. The phenomenological and hybrid models are in agreement in the high frequency regime, but diverge at low frequencies. The discrepancy implies that the attenuation of the GWB at low frequencies is not caused by energy being transferred into the environment through loss-cone scattering alone.}
    \label{fig:phenom-dadt-comp}
\end{figure}

The normalization $H_a$ is determined by setting a target lifetime via
\begin{equation}
    \tau_f=\int^{a_\mathrm{isco}}_{a_\mathrm{init}}\left(\left. \frac{da}{dt}\right|_\mathrm{ph}+\left.\frac{da}{dt}\right|_{\mathrm{gw}}\right)^{-1}da.
    \label{eq: tauf}
\end{equation}
The characteristic separation $a_c$ is the separation at which loss-cone scattering dominates dynamical friction---a quantity that our model computes for each binary. The phenom model assumes a constant value $a_c=100 \ \mathrm{pc}.$
The parameter $\nu_\mathrm{in}$ sets the slope of the hardening rate during the stellar-scattering regime of evolution, but also folds into it effects from the binary coupling to gas in the environment---something our model does not explicitly take into account. Evolution caused by dynamical friction is captured in the $\nu_\mathrm{out}$ parameter. Together the parameters $\nu_\mathrm{in}$ and $\nu_\mathrm{out}$ model interactions with the environment, which impact shape of the GWB spectrum \citep{2013CQGra..30v4014S}, with stellar- and gas-driven hardening causing a flattening/turnover at low frequencies \citep{2011MNRAS.411.1467K}. Figure \ref{fig:phenom-dadt-comp} shows the hardening time versus binary separation for both our core scouring model and the phenomenological hardening model plus gravitational wave emission. Our model predicts that SBHBs will decouple from their environment at larger separations that the phenom model. For this work we find GW emission will be the dominate hardening mechanism at separations of $\sim 10^{-1} \ \mathrm{pc}$. The overall effect of a larger decoupling radius between loss-cone scattering and GW emission is to increase the amplitude of the GWB spectrum at lower frequencies. 

To compare our results to those derived from the NANOGrav 15 year data set we use the SBHB population synthesis package \texttt{holodeck} \citep{2023ApJ...952L..37A} to generate GWB signals from mock SBHB populations. The package uses a semi-analytic model (SAM) submodule to define a grid in parameter space of total black hole mass, mass ratio, redshift, and binary separation/orbital frequency, with the GSMF, GPF, GMT, $M_\mathrm{BH}$--$M$ relation as inputs. The distribution of comoving number-density of cosmic SBHBs is then computed. GWB spectra are generated by providing the SAMs with values for Eq.\ (\ref{eq:phenom}). To obtain the appropriate values we fit the phenom model to the median hardening timescale generated by our model when considering $\hat H=1.6$ and $\langle t_c\rangle=1.4 \ \mathrm{Gyr}$. We find $ \nu_\mathrm{in}=-1.26, \ \nu_\mathrm{out}=2.53$, and $a_c=86 \ \mathrm{pc}$. The GWB generated using parameters informed by observed mass deficits over 1000 realizations shows that stellar hardening cannot be the sole source of the low frequency turnover observed in the GWB spectrum (see Fig. \ref{fig:phenom-dadt-comp}). To fully capture the turnover shown in the PTA data, contributions from other hardening mechansims such as eccentric SBHB orbits, gas-driven hardening, triple SBH systems, and SBH recoil will need to be considered. This is consistent with what was originally suggested by \citet{2025ACP...0204...01M}---the turnover in the GWB occurs at such small binary separations it is unlikely that interactions with loss-cone stars are the cause.
\section{Discussion}
\label{s:discuss}
We have found that the observed distribution of core sizes predicts an SBHB hardening rate that does not suppress the low frequency GWB spectrum (Fig. \ref{fig:phenom-dadt-comp}). Another way of stating this is that the GWB spectrum predicts stellar cores that are smaller for fixed galaxy stellar mass than are observed (Fig. \ref{fig:parmcomp}).

We have made a number of simplifying and practical assumptions in the course of our analysis which warrant discussion. We have assumed that (1) gas-dynamics play a negligible role in binary evolution for gas-poor mergers, (2) all SBHB orbits are circular throughout the entirety of their evolution, (3) the loss-cone is always full, (4) recoiling SBHs do not further enhance cores, (5) triple SBH systems do not occur, (6) binaries which have not coalesced by $z=0$ do not form a partial core, and (7) the $M_\mathrm{BH}$--$M$ relation is that of \citet{2013ARA&A..51..511K}, does not evolve with redshift, and applies to BCGs. Each of these assumptions affects the efficiency parameter $\hat H$ and variance of the model in different ways, which we discuss in the following subsections.
\subsection{Gas Dynamics}
Typically it is assumed in core scouring models that gas dynamics have a negligible effect on the resulting mass deficits, since core galaxies are formed from gas-poor mergers. It is possible that a galaxy could be gas-poor in the sense that there is not a sufficient reservoir of gas to trigger a nuclear starburst, and thus a scoured core would not be erased, but enough gas to interact dynamically with the binary. If the definition of gas-poor is only that there is not a reservoir of cold molecular gas that can be consumed in nuclear star-formation, then gas dynamics could play a role in binary hardening. An example of gas processes occurring in the center of core galaxies is M87, and it has long been known that a correlation exists between cores and radio-loud active nuclei \citep{2005A&A...440...73C, 2006A&A...447...97B}. Indeed, the fact that we require a more efficient hardening rate during the stellar-scattering dominated portion of the binary evolution could be, in part, explained by gas-driven hardening. If so it would imply that many SBHBs in core-galaxies will spend time as a binary AGN. The interaction of a binary with it's gaseous environment, the formation of circumbinary accretion disks, and the effects of such gas-driven evolution on the GWB have been the topics of numerous studies \citep[e.g.,][]{2011MNRAS.411.1467K, 2012MNRAS.427.2680K, 2013MNRAS.436.2997D,2020MNRAS.498..537S, 2023MNRAS.522.2707S, 2024MNRAS.534.2609S, 2025ApJ...994..168C}. However, gas-driven evolution of SBHBs in massive ellipticals implied by the analysis presented here cannot be responsible for the low frequency turnover observed in the GWB. Figure \ref{fig:phenom-dadt-comp} shows that even when considering $\hat H=1.6$ we do not recover the hardening rates obtained by the NANOGrav 15 year analysis.

We have made the argument that SBHBs responsible for core galaxies will be the same SBHBs that dominate the GWB signal ($M_\mathrm{bin}\gtrsim 10^9 M_\odot$ and $q\gtrsim 0.1$, \citealt{2017MNRAS.464.3131K}). One possible explanation is that SBHBs formed in gas-rich mergers may play a comparable role, making power-law (cusp) galaxies a target of investigation. While cores dominate at $\mathcal{M}_B \lesssim18$ there are still massive, luminous power-law galaxies which would have hosted SBHBs. In fact, recent studies have confirmed ultramassive black holes (UBHs) at the centers of massive, core-less galaxies \citep{2025arXiv251204178D}. A possible explanation is that these were galaxies whose cores were erased in a major, gas-rich merger. Further, if gas-dynamics does tend to drive $q$ to unity, these binaries could have a reasonably large impact on the GWB. Large star-forming spirals seem to be unlikely candidates, as BH-pairs evolving in clumpy, gas-rich galaxies (ULIRGS or Starburst galaxies in $z\sim1$--$3$) undergo stochastic decay, scattering from GMCs and clusters. Ultimately this causes a random walk and stalls the binary at the $100$--$1000 \ \mathrm{pc}$ scale \citep{2017MNRAS.464.2952T}. Thus the inclusion of power-law ETGs and the incorporation of gas-dynamics is an important next addition to our model if we are to fully disentangle the effects of gas dynamics with those of stellar scattering.
\subsection{Eccentricity}
\label{ss: ecc}
We have assumed all SBHB orbits are circular and remain circular. However, the prediction of high initial eccentricities for SBHB orbits is common in hydrodynamical simulations of galaxy mergers and N-body experiments predict eccentricity evolution \citep{2011ApJ...732L..26P,2012ApJ...749..147K,2022MNRAS.511.4753G,2024MNRAS.532..295F, 2023MNRAS.526.2688R, 2025A&A...703A..86F, 2026MNRAS.546ag026G}. In general, each of the hardening mechanisms affect and are affected by eccentricity, leading to changes in the characteristic strain spectrum of the nanohertz GWB. During the loss-cone phase of binary hardening, three-body interactions with stars will tend to increase the orbital eccentricity \citep{1996NewA....1...35Q, 2006ApJ...651..392S}, while GW emission circularizes orbits \citep{1964PhRv..136.1224P}. The mass ejection rate has only a weak dependence on the eccentricity; however, the hardening rate for stellar-scattering \textit{increases} for more eccentric orbits, rising by $\sim 10\%$ between eccentricities of 0.0 and 0.9 \citep{2006ApJ...651..392S}. \citet{2017MNRAS.471.4508K} find that for massive binaries with large mass ratios ($M_\mathrm{bin}\gtrsim 10^9 M_\odot$ and $q\gtrsim 0.1$) eccentricities monotonically decrease approaching the PTA-sensitive band. In the most eccentric models, with initial eccentricity $e_0=0.95$, by the time the binary has reached PTA band separations the eccentricity has fallen to $e\sim0.5$.
When including the effects of a circumbinary disk, \citet{2023MNRAS.522.2707S} found that there is an equilibrium eccentricity ($e\sim0.5$) all binaries with $0.1 \leq q \leq1.0$ tend toward. These results are consistent with analyses of PTA data which suggest a mild eccentricity in the SBHB population, the signature of which is a flatter or inverted spectrum at lower GW frequencies \citep[e.g.,][]{2017PhRvL.118r1102T}. The \citet{2024A&A...685A..94E} found models with high central stellar densities and high orbital eccentricities are favored. It seems reasonable, then, that our best-fit value of $\hat H$ above unity could be explained predominately by eccentric orbits, as this would affect the hardening timescale for all binaries in the model. Even so, if eccentric orbital evolution during the stellar scattering regime explains the hardening rates found in our model, the effect on its own does not explain the discrepancy with the NANOGrav results. Additional modeling of eccentric orbits will be needed in order to break its degeneracy with hardening from other sources such as gas dynamics and prompt mergers from triple interactions.
\subsection{Loss-cone Depletion}
The rate at which a binary hardens by stellar scattering is governed by the rate at which the loss-cone is repopulated. It has long been noted that a purely spherical potential does not allow for the loss-cone to be refilled faster than the relaxation timescale, causing SBHBs within the host galaxy to stall \citep{2003AIPC..686..201M, 2006ApJ...648..976M}. It has also been shown that even mild triaxiality in the potential results in a full loss-cone because of the existence of box orbits not seen in spherical potentials \citep{2002MNRAS.331..935Y,2006astro.ph..1520H, 2011ApJ...732...89K, 2017MNRAS.464.2301G,2021ApJ...922...40D}. We have, therefore, assumed a full loss-cone for each binary. The fact that our model prefers a more efficient hardening rate despite a full loss-cone gives confidence that the assumption is appropriate, though future work will seek to include triaxial potentials and a variable loss-cone.
\subsection{Supermassive Black Hole Recoil}
\label{ss: recoil}
During the final stages of coalescence GW emission from a black hole binary occurs anisotropically, resulting in a kick to the remnant black hole \citep{1973ApJ...183..657B}. The velocity of the kick depends on both the mass ratio and spin alignment of the black hole pair, with kick velocity depending non-linearly on mass ratio, spin magnitude, and spin orientation. Kicks range from a few hundred kilometers per second to a maximum value of $\sim$4000 km/s \citep{2007ApJ...659L...5C}. Numerous studies have shown that recoiled SBHs in the nuclei of gas-poor remnants enhance core size/mass beyond that induced by binary activity by a factor of $2$--$3$ and that the enchanced mass deficit is an increasing function of kick velocity \citep{2004ApJ...613L..37B,2008ApJ...678..780G,2021MNRAS.502.4794N,2024MNRAS.532.4681P,2024ApJ...974..204K,2025MNRAS.537.3421R}. This adds a complication to using mass deficits as a probe into SBHB hardening timescales. If we expect that all massive early-type galaxies with cores experienced a recoiling SBH then all observed mass deficits are a combination of mass scoured out from binary activity and mass removed the recoiled SBH. This means that stellar scattering does not need to explain all of the mass deficit, resulting in an increased value of $\hat H$ in our model, all else being equal. If, on the other hand, mass deficit enhancement by recoil is not common we do not expect the overall amplitude of the $M_\mathrm{def}$--$M_\star$ relation to be affected, but instead would explain some of the observed variance (see Fig. \ref{fig:fN}). Indeed, observations suggest that the core-galaxies NGC 4486B, Abell 2261 BCG, and Abell 402 BCG were hosts to recoiled SBHs, and each of these galaxies have large cores for their stellar masses \citep{2012ApJ...756..159P,2025arXiv251214695T, 2026arXiv260310104M}. 

It is interesting to consider the combined effect of SBH recoil and eccentric SBHB orbits. The effect of eccentric orbits in the binary population is to, in general, decrease the mass ejected in stars by the binary while increasing the hardening rate. On the other hand, recoiled SBHs \emph{increase} the observed mass deficit. If the mass deficits in Figure \ref{fig:Obs sample} are in general characterized by $M_\mathrm{def}=M_\mathrm{ej}^\mathrm{binary}+M_\mathrm{ej}^\mathrm{recoil}$ then our model underestimates $\hat H$ and thus overestimates the median hardening timescale. Since recoil and eccentricity tend to counteract each other modeling both will be crucial for understanding the tension between the derived hardening timecales here and those implied by the nanohertz GWB.
\subsection{Triple Supermassive Black Holes}
If the SBHB coalescence timescale is long compared to the galaxy merger timescale then triple SBH systems may be common \citep{2018MNRAS.477.2599B}. These triplets are also a natural way to decrease the hardening timescale for binaries while at the same time scouring out large cores through tertiary ejection. \cite{2024MNRAS.527.7424S} find in the Illustris-1 simulation that 22\% of binaries formed using their prescriptions resulted in a triple system, where a triple was defined as an instance where an intruder SBH overtakes a binary. Of these systems more, than 70\% would not have coalesced by $z=0$ otherwise, and $\sim6\%$ form strong triplets at parsec-scale separations. Work by \cite{2025ApJ...993..222S} showed that of the strong triple systems 21\% led to prompt coalescence of the original binary, 48\% led to coalescence after a slingshot kick to the lightest SBH, and 31\% resulted in no coalescence.

In this work we define a triplet as any time an SBHB host experiences a major merger before the binary coalesces. When this happens we ignore the system in the model. For the 27203 TNG-300 and TNG-Cluster subhalos in our sample this occurs roughly 20\% of the time for our best-fit value of $\hat H$. We have assumed that the frequency of strong triplets is small enough that they play a negligible role in core formation at the population level. This may not be true for the most massive cores which have undergone multiple major mergers. If this assumption is incorrect there are multiple effects to consider. If a triplet is gravitationally bound there will be core scouring from the formation of the binary and additional scouring from the inbound third black hole. Since triplet systems tend to eject the tertiary SBH one might expect an enhanced stellar mass deficit depending on the kick (see \ref{ss: recoil}). However, this is complicated by the fact that a prompt merger of the binary can ``skip'' a portion of the stellar scattering evolution. Thus, further investigation is required to make a definitive statement of the effect of triple systems on core scouring.
\subsection{Partial Core Scouring}
It is possible that an SBHB has stalled (i.e., has not coalesced) by $z=0$ and is located in the stellar-scattering regime of its evolution. In that case a partial core would be scoured, presenting itself as a smaller mass deficit compared to other core galaxies at the same stellar mass. This phenomenon is unlikely to have occurred in any of the observed core galaxies, since none of them host an SBHB candidate. Therefore, we do not include partial scouring in our model. We also exclude black hole pairs which occur in the model that have not coalesced by $z=0$. A fraction of these pairs have final separations at the parsec scale, within the hard-binary radius. These binaries could ostensibly have very large cores, and their exclusion could bias our results, though the number of these binaries is always small compared to the rest of the cored population generated by the model.
\subsection{Scaling Relations}
It is an empirical fact that, at $z\sim0$ the mass of a galaxy's central black hole is related to the mass/luminosity of the same galaxy's central bulge and to its central dispersion \citep{2009ApJ...698..198G,2013ARA&A..51..511K}, though the details of these relationships are uncertain at the highest masses/luminosities \citep{2013ARA&A..51..511K, 2013ApJ...764..184M, 2016MNRAS.460.3119S}.  Additionally, the mass of the central SBH from scaling relations for brightest cluster galaxies may be unreliable (differences between say $M$--$L$ and $M$--$\sigma$ \citep{2007ApJ...662..808L, 2025arXiv251204178D}
In this work we adopt the $M_\mathrm{bh}$--$M_\mathrm{bulge}$ relation of \citep{2013ARA&A..51..511K}, which has a higher amplitude and steeper slope than other relations. Since $M_\mathrm{def}\propto M_\mathrm{bin}$ our results are sensitive to our choice of scaling relation. We note here that using a scaling relation that produces lower mass black holes will tend to drive the median posterior value of $\hat H$ closer to unity and increasing coalescence times. The main result that stellar hardening with circular orbits alone cannot explain the nanohertz gravitational wave background is unaltered by our choice of black hole mass scaling relation.

James Webb Space Telescope (JWST) observations of over-massive SBHs and dual AGN at large redshifts, challenge black hole--galaxy co-evolution models \citep{2023ApJ...959...39H,2023ApJ...957L...3P,2024ApJ...963..129M, 2024A&A...691A.145M}. A solution to this, and to the larger than expected amplitude of the GWB \citep{2023ApJ...951L...8A, 2023A&A...678A..50E, 2023ApJ...951L...6R, 2023RAA....23g5024X} is an evolving $M_\mathrm{BH}$--$M$ and/or $M_\mathrm{BH}$--$\sigma$ relation \citep{2007ApJ...670..249L,2023MNRAS.524.4403M,2008ApJ...676...33D, 2023ApJ...949L..24S,2024ApJ...974..261C,2025arXiv250818126M, 2025ApJ...988...90H}. If there is significant evolution in the redshift range $0 < z \lesssim 2$ such that SBHs are more massive than predicted by the local relations, we would expect that to be reflected in the size of cores, since $M_\mathrm{def}\propto M_\mathrm{bin}$. For the same reason a general increase in the amplitude of the $M_\mathrm{BH}$--$M$ relation would also increase the size of cores. The overall increase in masses would also impact the total hardening timescales, as all three regimes considered in this work depend strongly on the binary mass. It is possible that including an evolving black hole mass scaling relation would demand a higher value of $\hat H$ to explain cores, thereby alleviating some of the tension highlighted in this work, but not all. Alternatively, if the amplitude of the GWB is mostly explained by a top-heavy GSMF \citep{2024ApJ...971L..29L} we would expect the size of cores to be unaffected. Future work including an evolving $M_\mathrm{BH}$--$M$ relation to the core scouring model presented here would provide important constraints on the contributions of each of these effects to the amplitude of the GWB.
\section{Summary}
\label{s: sum}
We have made progress in constraining the properties of the SBHB population by using both EM and GW datasets. In section \ref{s: Methods} we presented a general method for estimating stellar mass deficits in local core galaxies using the Nuker law. We also develop a model of core formation which predicts stellar mass deficits. In our post-processing model of core scouring, we use observation- and simulation-informed prescriptions to determine whether core scouring occurs within a given galaxy hosting an SBH pair, and how much mass is scoured out. Section \ref{s: results} shows that our model can only produce mass deficits consistent with the observed population if the hardening timescale is shorter than predicted by N-body results by a factor of about 1.6, resulting in longer coalescence times than predicted by models informed by the GWB (Fig. \ref{fig:merge_times}), though consistent with other semi-analytic models. Additionally, we find that the stellar hardening timescales found in our best-fit model, when applied to the NANOGrav 15 year dataset, do not reproduce the attenuation found in the GWB at low frequencies (Fig. \ref{fig:phenom-dadt-comp}). Neither do populations generated using \texttt{holodeck} based on the 15 year NANOGrav data reproduce the observed mass deficits in core galaxies. Taken together, our results identify a tension between the SBHB hardening rates inferred from local core-galaxy mass deficits and those required to reproduce the apparent low-frequency attenuation in the nanohertz GWB. However, this tension should not be interpreted as uniquely implicating any single physical mechanism. Several effects not included in our fiducial model—including eccentric binary evolution, gas-driven hardening, gravitational-wave recoil, triple-SBH interactions, incomplete loss-cone refilling, uncertainties in the adopted $M_{\rm BH}$--galaxy scaling relation, and contributions from power-law or gas-rich galaxies—can alter both the inferred mass deficits and the GWB spectrum. Our analysis therefore demonstrates that core-galaxy constraints provide an important, complementary probe of SBHB evolution, but a more complete population model incorporating these effects is required to determine whether the apparent GWB turnover reflects environmental coupling, highly eccentric orbits, population demographics, or some combination thereof.
\section{Acknowledgments}
The authors would like to thank Chiara Mingarelli for useful and thought-provoking comments as well as Joseph Simon and Luke Zoltan Kelley for insightful discussions on binary hardening mechanisms. We would also like to thank the reviewer for their valuable comments which have greatly improved the paper.

\section{Data Availability}
The data underlying this work are deposited in Deep Blue Data, the University of Michigan's institutional data repository. The data will be publicly accessible and maintained for a minimum of ten years and can be accessed via https://doi.org/10.7302/77h9-b552.
\bibliography{references}{}
\bibliographystyle{aasjournalv7}
\end{document}